\documentclass[
 aps,
 pre,
 reprint,
 amsmath,amssymb,
 showpacs,
 floatfix
]{revtex4-1}
\usepackage{graphicx}
\usepackage{dcolumn}
\usepackage{bm}
\usepackage{epsfig}
\begin{document}
\preprint{APS/123-QED}
\title{ Drastic Changes in Dielectric Function of Silver Under dc Voltage }
\author{B.V. Kryzhanovsky}
 \email{kryzhanov@gmail.com}
\author{ A.N. Palagushkin}
\author{ S.A. Prokopenko }
\author{ A.P. Sergeev }
\affiliation{ Department of Nanotechnologies, Center of Optical Neural Technologies SRISA RAS, Moscow}
\author{ A.O. Melikyan } 
 \email{armen{\_}melikyan@hotmail.com}
\affiliation{ Department of Physics, Russian-Armenian (Slavonic) Univercity, Yerevan}

\date{\today}
\begin{abstract}
Significant changes of the relative permittivity of a silver film have been detected using the surface plasmon resonance (SPR) method when a constant electric field is applied to a MDM (metal-dielectric-metal) nanostructure. The structure looks like a capacitor with a $177 - nm$ dielectric corundum film placed between two silver films $49nm$ and $36nm$ thick. The effect manifests itself as a noticeable change of the reflectivity of the structure when the voltage of up to $30V$is applied to the electrodes. We have a good agreement between the theory and experiment only if we suppose that the optical parameters of anode and cathode silver films change differently and the $A{l_2}{O_3}$ film absorbs the incident light. The refraction coefficient of the cathode silver layer is shown to become zero when the applied voltage is above $16V$.
\end{abstract}
\maketitle

\section{Introduction}
The optical properties of thin metal films secure their wide use in different fields – from mirrors to supersensitive detectors \cite{Maier} capable of detecting refraction coefficient changes as small as ${10^{ - 7}}$ \cite{Mikaelian}. These properties are considered to be stable to external actions and changes of the absorption spectrum are detected only in going from macro samples to nano samples. The latter effect is caused by collective oscillations of electrons (so-called plasmon oscillations) in nanoparticles  with  sizes smaller than  the electron mean free  path  in massive samples. It was revealed that the static polarizability of metal clusters containing  several tens of atoms, which determines the frequency of collective oscillations, differs sufficiently from  macroscopic polarizability \cite{Knight}. Besides, calculations show that in $Ag$ clusters $1 - 2 nm$ in diameter an extra charge of $1$ to $3$ electron charges should shift the plasmon absorption line by about $5 nm$, while for even $10 - nm$ clusters this spectral shift is negligibly small \cite{Quinten}. It sounds reasonable because the spectral shift is caused by a change of the charge density, which is noticeable in small clusters and negligibly small in large clusters. It is however known that plasmon frequencies are strongly dependent on the form and size of nanoparticles. The examples show that noticeable changes of optical properties of metal nano-size samples are hard to cause without changing their geometry.

	In the present communication we give the experimental data that clearly show a considerable change of the effective relative permittivity of $Ag$ layers in the MDM nanostructure.

\section{ Experiment}
\begin{figure}
\includegraphics[width=8.6cm]{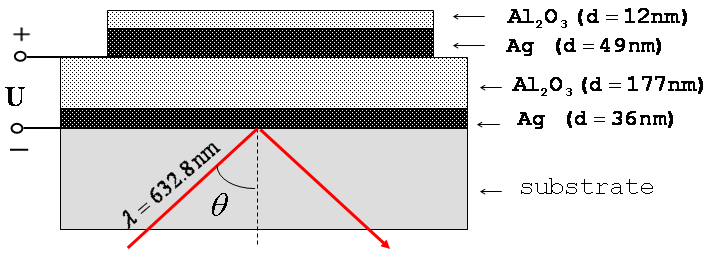}
\caption{\label{fig:fig01}MDM plasmonic structure}
\end{figure}

High-vacuum electron-beam deposition was used to make plasmon 4-nanolayer MDM samples ($A{l_2}{O_3}/Ag/A{l_2}{O_3}/Ag$). Figure~\ref{fig:fig01} illustrates a 4-layer nanostructure whose optical properties are studied in the experiment. The layers of the structure are $12,\,49,\,177\;{\textrm{and}}\;36\,nm$ thick. A quartz substrate was kept at a relatively low temperature (about ${150^ \circ }C$) in the process of layer deposition which allows obtaining smaller crystal grains in the layers  and was necessary to ensure the surface smothness of the film. The outer $A{l_2}{O_3}$ layer ($d = 12nm$) serves as antioxidation coating. No adhesive sublayers were used. A laser with wavelength $\lambda  = 632.8nm$ and $p$-polarization was used to generate SPR. It should be noted that this structure can be considered as a metal-dielectric plasmonic waveguide with two SPR angles ${\theta _1} = {43^0}$ and ${\theta _2} = {56.3^0}$.

	In deposition we used masks to form a thin-film capacitor structure. The diameter of the outer silver electrode was 3 mm. A constant electric field was applied to the capacitor. The structures could stand up to $30V$ without electric breakdown. This corresponds to the field strength of $1.7 \cdot {10^6}V/cm$ in the dielectric layer, given the corundum permittivity $\varepsilon  = 10$. However, when the air humidity was high, a lower breakdown voltage and greater leakage currents were observed.

	In the experiment we  measured the SPR spectrum (the angular reflection spectrum of the MDM structure) using the Kretschmann configuration \cite{Kretschmann}. We used an automatic semicylinder $\theta - 2\theta $ goniometer with additional plane-wavefront-forming optics to measure the angular spectrum of total reflection.

\begin{figure}
\includegraphics[width=8.6cm]{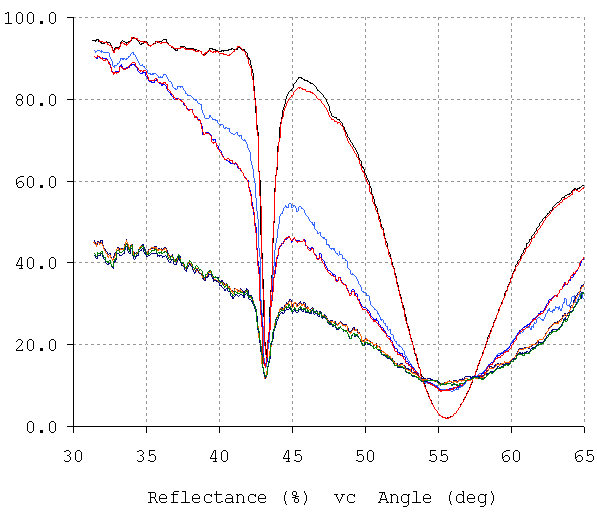}
\caption{\label{fig:fig02}The reflection spectrum versus the voltage $U$ across the MDM structure. The upper curves corresponds to $U$changing from $0 V$ to $28 V$ in $4 - V$ steps. The lowest curve corresponds to near-breakdown voltage $U = 30 V$.}
\end{figure}

	The experimental results are given in Figure~\ref{fig:fig02} which shows how the reflection spectrum changes with the electric field changing from $0$ to $30$ volts. The curves allow us to assert that the reflection coefficient changes noticeably when the voltage increases above $4V$, at $U > 16V$ it reaches  saturation. The resonance peak width grows considerably with $U$, yet the shift from the resonance angles ${\theta _1} = {43^0}$ and ${\theta _2} = {56.3^0}$ does not occur. The reflection coefficient changes significantly: when $U$ approaches $20 V$, it decreases fourfold near the first resonance angle and increases fivefold near the second SPR angle. In Figure~\ref{fig:fig02} we can select three groups of curves: the first group corresponds to $U \in [0,\;6V]$, the second to $U \in [8 V,\;16 V]$, and the third to $U \in [18 V,\;32 V]$. The grouping will become clear in the next paragraph where we examine the voltage dependence of optical parameters. Besides, there are two points where the reflection coefficient is independent of the applied voltage: $R \approx 11.6\% $ at ${\theta} = {53.95^0}$ and  $R \approx 11.8\% $ at  ${\theta} = {57.39^0}$ for any $U$. It is hard to explain the existence of these points, however they can be used for testing the results: the theoretical curves should intersect at the points.

	The curves of reflection coefficient $R$  versus angle of incidence $\theta $ allowed us to compute the optical parameters of all layers of the MDM structure (optical thickness, refraction coefficients $n$ and absorption coefficients $k$) as function of voltage $U \in [0,\;30 V]$. The effective optical parameters of the layers were determined by the best agreement between experimental and theoretical curves.

\begin{figure}
\includegraphics[width=8.6cm]{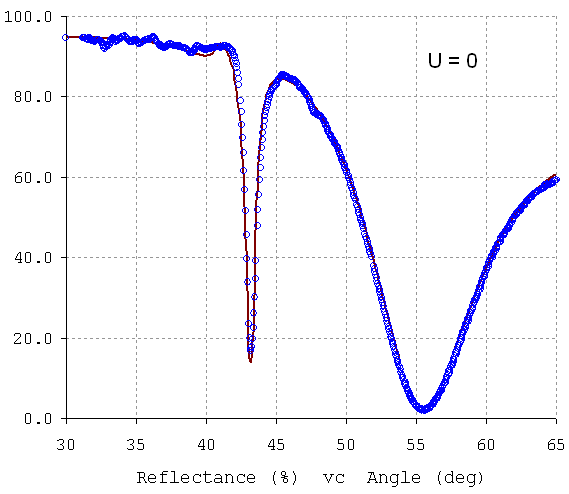}
\caption{\label{fig:fig03}SPR spectrum when there is no electric field (the solid line corresponds to the theory, marks to the experiments).}
\end{figure}

	The conventional methods and software were used in computation of theoretical angular spectrum (the method is described in \cite{Palagushkin1,Palagushkin2} in detail). The optical constants of silver and corundum from the SOPRA database and layer thicknesses measured during the deposition process were taken as initial values for the theoretical model in the absence of the external electric field. Then we computed the best agreement between the theoretical and experimental curves $R = R(\theta )$ to correct the layer thicknesses to use them for further computations. The results of simulation are shown in Figure~\ref{fig:fig03} and in Table~\ref{table:table1} for $U = 0$.

\begin{table}
\caption{\label{table:table1} The initial optical parameters of the layers ($U=0$)}
\begin{ruledtabular}
\begin{tabular}{ccccc}
No & material 		& $n$ 	& $k$ 	& thickness, ($nm$) \\
1  & ${Al_2}{O_3}$	& 1.659	& 0		& 12.02\\
2  & $Ag$			& 0.136	& 4.011	& 49.20\\
3  & ${Al_2}{O_3}$	& 1.659	& 0		& 177.36\\
4  & $Ag$			& 0.136	& 4.011	& 36.46\\
5  & substrate		& 1.525	& 0		& -\\
\end{tabular}
\end{ruledtabular}
\end{table}

	Given non-zero external electric field ($U > 0$), the thicknesses of the layers were considered fixed (Table~\ref{table:table1}) in computations and effective optical parameters of the $Ag$ and $A{l_2}{O_3}$ layers (refraction $n$ and absorption $k$ coefficients) were varied to reach the best agreement with the experimental spectrum. All parameters of the outer protective layer ($d = 12nm$) were also fixed according to Table~\ref{table:table1} because it should have no electric field in it.

\begin{figure*}
\includegraphics{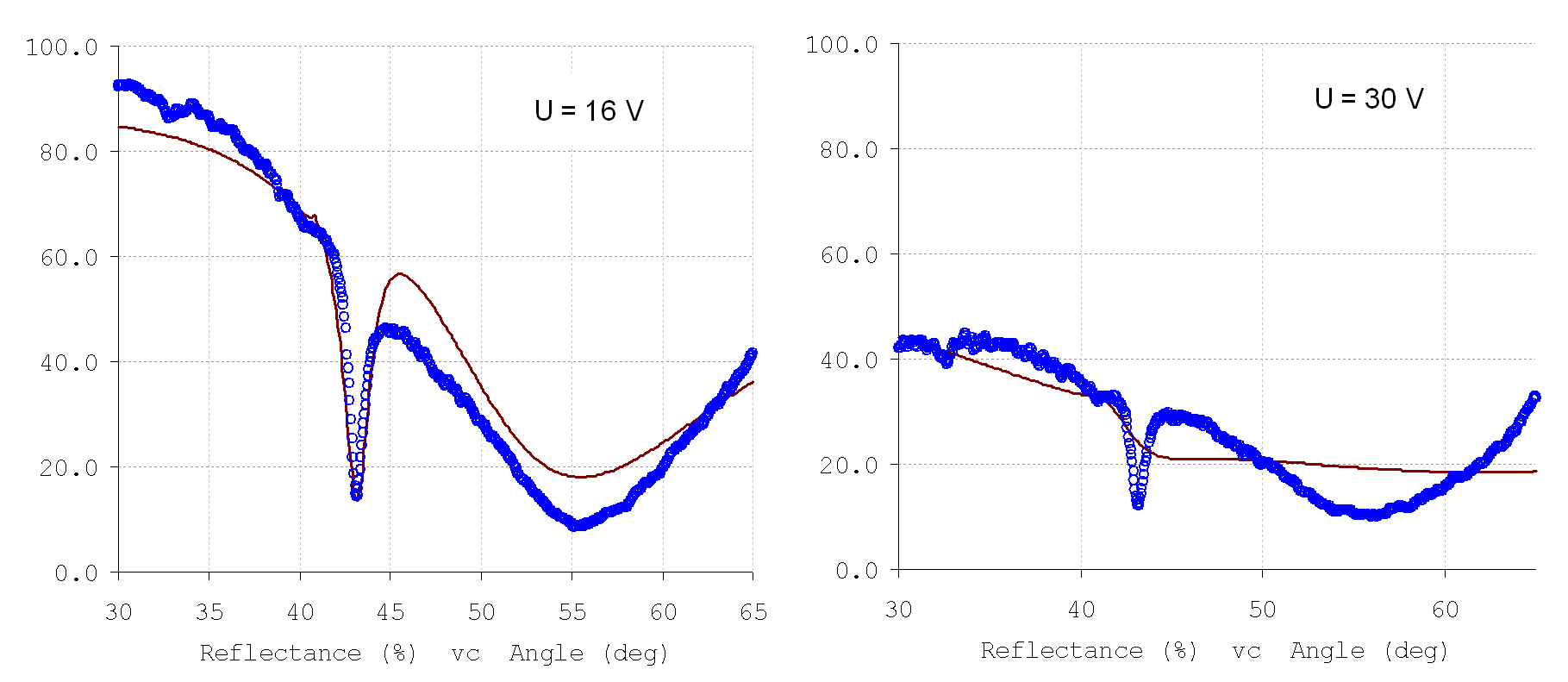}
\caption{\label{fig:fig04}SPR spectra for $U = 16 V$ and $U = 30 V$ (the marks are the experimental data, the solid line are the computation results). The theoretic curve  are built under assumption that the optical parameters of the both $Ag$ layers are the same. It is seen that these curves do not agree well with the experiment.}
\end{figure*}

	Comparing the experimental and theoretical curves allows the following conclusions. With no external field (Fig.~\ref{fig:fig03}) the model is nearly perfect: the original values of optical parameters of the layers are identical to those given in the SOPRA data base. If there is an external field, the simulation needs closer consideration. If we assume that the optical properties of both $Ag$ layers keep the same in the presence of the electric field, the difference between theoretical and experimental data will increase with the field strength and we can’t eliminate it by varying $n$ and $k$. Indeed, the theoretical curves in Fig.~\ref{fig:fig04} built from this assumption agrees with the experiment poorly. It is clearly caused by the fact that the computations do not take into account nonlinear effects which can take place when the electric field strength is high. Surface charges caused by dielectric polarization and changes of  the numbers of electrons  in the metal layers should also be taken into consideration. The theory and experiment agree better if we assume that the properties of $Ag$ of the cathode and anode differ dramatically. Validation of such approach and corresponding simulation results are given in the next section.

\section{ The results of the simulation}
The difference between the theory and experiment can be eliminated if we suppose that the optical parameters of the upper (anode) and lower (cathode) $Ag$ layers can be different and must be optimized  separately. The separate optimization is possible under the assumption that charges accumulate at the dielectric-metal interface and the electron plasma density changes across the metal layers in a strong electric field.

	Indeed, when the dielectric layer is about $d_0\sim200\,nm$ thick and the applied voltage is nearly $30\,V$, the surface charge density is $\sigma  = U/4\pi \varepsilon {d_0} \sim 40CGSE$ (the permittivity of corundum $\varepsilon \sim 10$). If the layer area is $1\;c{m^2}$, the number of excess electrons on one of the $Ag$ layer (and deficient ones on the other) is  $\sim{10^{11}}$, while the full number of conduction electrons is $\sim{10^{18}}$.

	Let us ask ourselves what the deficiency of electrons in the layer can result in. It is clear that the depth of the potential well for all electrons of the anode increases by $30\;eV$. At the same time, we should take into account the changed band structure. The matter is that the additional attraction force from the uncompensated positive charge brings about a shift of the conduction band and, of course, the Fermi level ${E_F}$ down to a lower energy. Correspondingly, this results in the interband  absorption  threshold  ($\sim4\,eV$)  lowering. However, the shift of the d-electron band is much smaller because of  stronger binding to the nuclei.

The change of the interband absorption level changes the dielectric permittivity of the metal and, therefore, parameters $n$ and $k$. For instance, the model function $\varepsilon (\omega )$  is built for gold (see \cite{Maier}) which allows for two interband transitions. The eleven  fitting parameters (resonance frequencies, resonance widths, complex amplitudes, asymptotic parameter, etc.) of the function allowed to reach a good agreement with the data \cite{Etchegoin}. In the case of silver a similar function has not been built so far, so we limit ourselves to the rough estimation using formulae for gold \cite{Maier}. A $0.1 - eV$ change of threshold $\Delta $ results in an about unit change of the dielectric permittivity, which can be detected experimentally. Note, that the formula presented  in \cite{Etchegoin2} describes well the permittivity of $Ag$ for the photon energies less than $3\,eV$, which does not allow to analyze the influence of  the shift of  the interband  transition threshold on the permittivity value at $\lambda  = 632.8nm$ ($1.96\,eV$).

	At the same time, the excess of electrons on the cathode raises the interband absorption level. Excess electrons do not leave the metal as long as the attraction caused by image charges exceeds the repulsion caused  by excess electrons. The condition of equilibrium of  these forces can be written as ${{e}^{2}}/{{l}^{2}}=eU/2d_0$, where $U/2d_0$ is the field strength on the cathode (it is twice greater in the capacitor itself), $l$ is the electron-cathode distance. Then the saturation voltage is ${U_s} = 6 \times {10^{ - 12}}/{l^2}$. Let us suppose that slow electrons are scattered by defects whose concentration is about ${10^{18}}\;c{m^{ - 3}}$. The average free path is about $10nm$ and saturation voltage is a few volts in this case. Further increase of $U$ leads to dynamic equilibrium, that is, the number of electrons arriving to the cathode in unit time is equal to the number of electrons leaking through the dielectric per unit time.

\begin{figure*}
\includegraphics{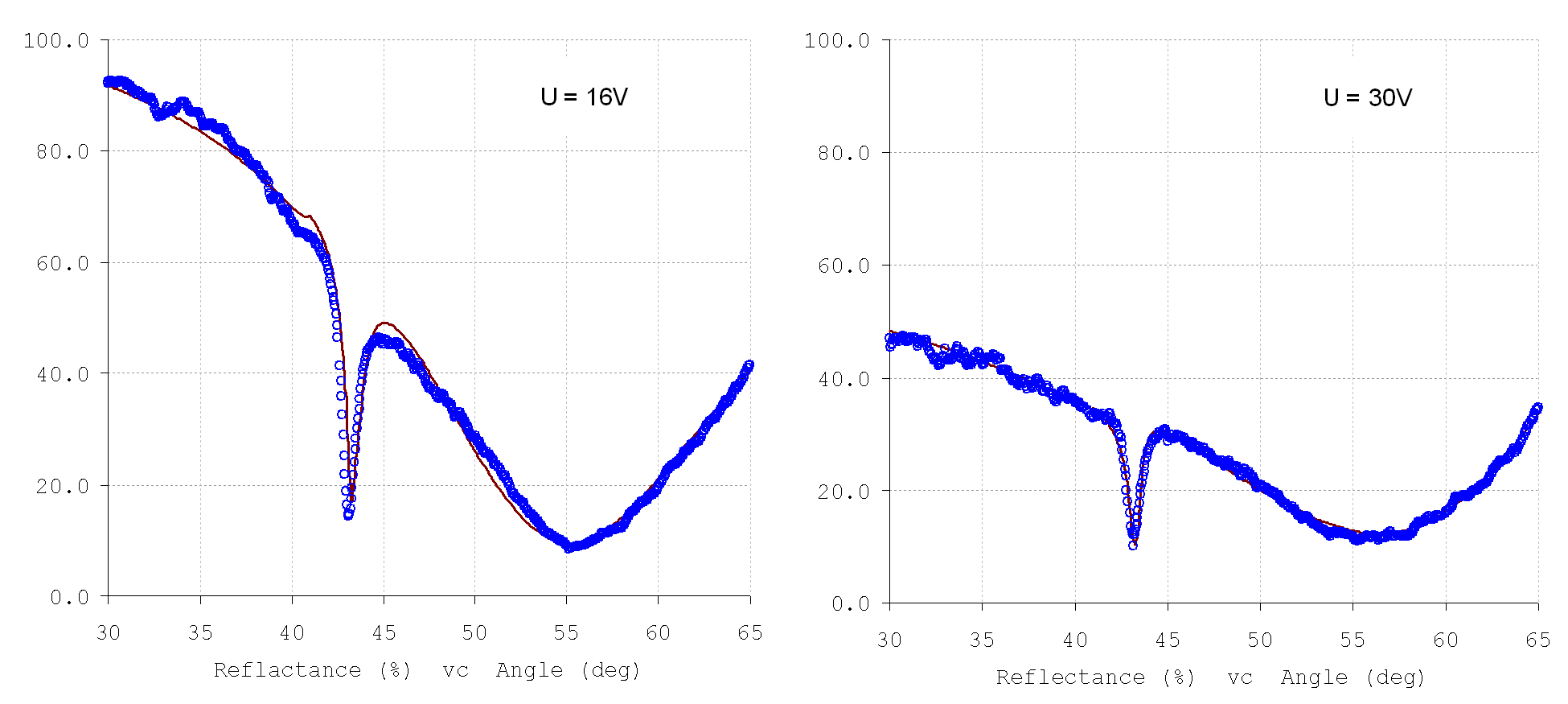}
\caption{\label{fig:fig05}SPR spectra for $U = 16 V$  and  $U = 30 V$ (marks – experiment, line – theory). The theoretic curves are built given separate optimizations of $Ag$  parameters on the anode and cathode.}
\end{figure*}

	The above considerations suggest the necessity to optimize optical parameters of the upper (anode) and lower (cathode) $Ag$ layer independently. Indeed, the modelling using independent optimization of anode and cathode layers gives good agreement between theoretic and experimental curves. Figure~\ref{fig:fig05} gives the results of simulation of an SPR angular spectrum for $U = 16\,V$ and $U = 30\,V$. Comparing Figure~\ref{fig:fig04} and Figure~\ref{fig:fig05} shows that the independent optimization allows a significantly better theory-experiment agreement. The computed values of optical properties for this kind of modelling are given in Table~\ref{table:table2}. The first thing that draws attention is that the refractive index of the anode layer proves to be smaller than that of the cathode. This is an expected observation which agrees with measurements made during the deposition process and is explained by the fact that the deposition on a smooth substrate always gives a better quality of the $Ag$ layer than in deposition on a relatively loose intermediate layer. Secondly, with the growing voltage the $A{l_2}{O_3}$ layer starts exhibiting slight absorption and its refractive index decreases. Probably, it is caused by electrons arriving to this layer from the cathode. To avoid misunderstanding, we should point out that in modelling it is impossible to take into account surface irregularities, interpenetration of layers and non-linear effects caused by a strong electric field. That is why the values of $n$ and $k$ given in Table~\ref{table:table2} should be regarded as certain effective values that allows for the drawbacks of this theoretical approach in a way.

\begin{table*}
\caption{\label{table:table2} The optical parameters of the MDM structure}
\begin{ruledtabular}
\begin{tabular}{ccccccc}
   & 
\multicolumn{2}{c}{$Ag$ (anode), $d=49 nm$} &
\multicolumn{2}{c}{${Al_2}{O_3}$, $d=177 nm$} &
\multicolumn{2}{c}{$Ag$ (cathode), $d=36 nm$} \\
$U(V)$ 	& $n$ 		& $k$ 	& $n$ 	& $k$ 	& $n$ 	& $k$ \\
0  		& 0.1360	&4.0110	&1.6908	&0.0000	&0.1341	&4.0100\\	
4  		& 0.1341	&4.0109	&1.6592	&0.0022	&0.1340	&4.0100\\
8  		& 0.5928	&4.2967	&1.6593	&0.0087	&0.0881	&3.4853\\
12  	& 0.5931	&4.2969	&1.6533	&0.0116	&0.0882	&3.4852\\
16  	& 0.5932	&4.2970	&1.6531	&0.0113	&0.0882	&3.4852\\
20  	& 0.3752	&4.6909	&1.5059	&0.0583	&0.0000	&1.0263\\
24  	& 0.3717	&4.6840	&1.5064	&0.0589	&0.0000	&1.0308\\
28  	& 0.3828	&4.6885	&1.5122	&0.0622	&0.0000	&1.0619\\
30  	& 0.3778	&4.7022	&1.5161	&0.0642	&0.0000	&1.0816\\
\end{tabular}
\end{ruledtabular}
\end{table*}

\begin{figure*}
\includegraphics{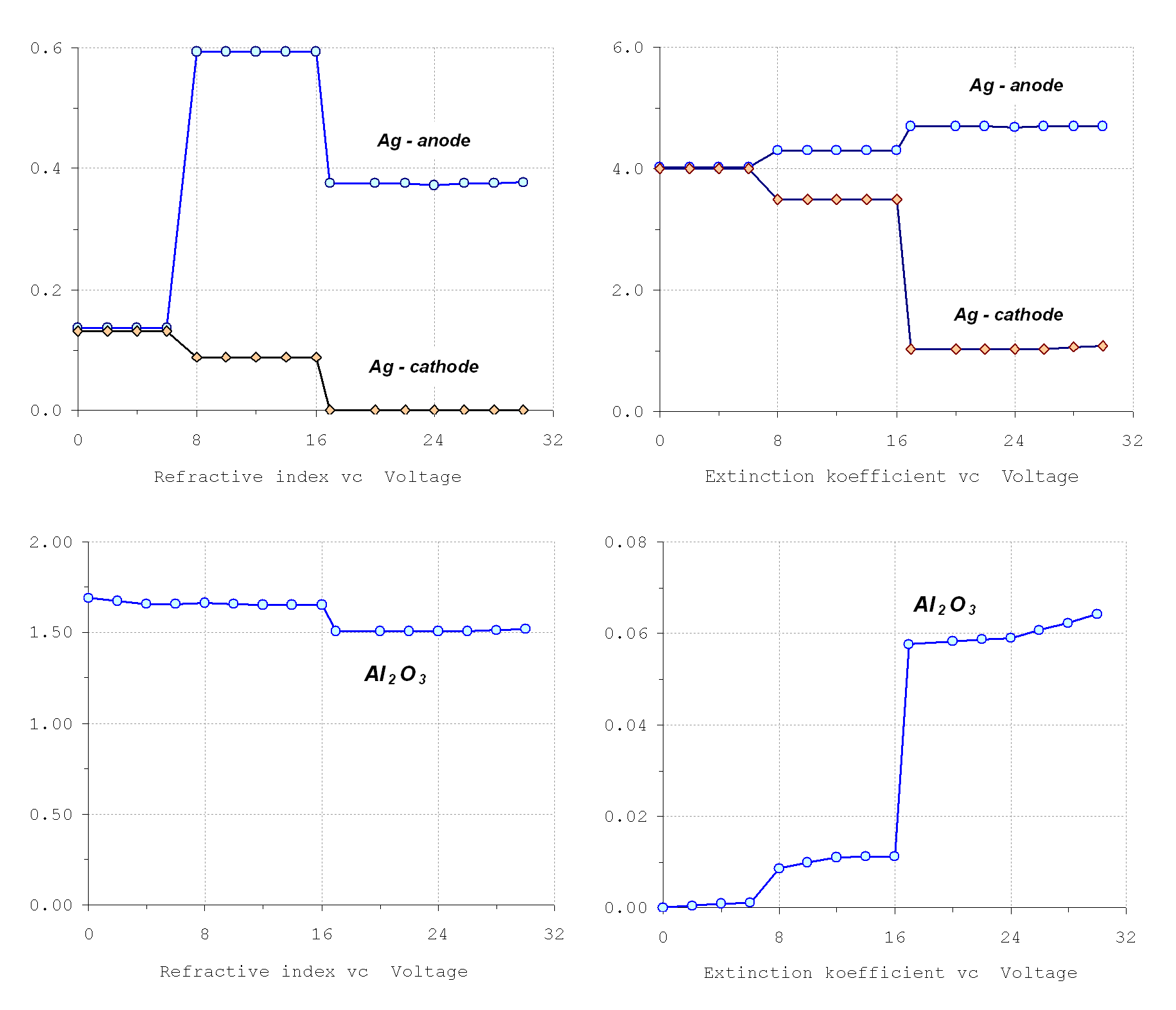}
\caption{\label{fig:fig06}The refractive indices and absorptions coefficients versus voltage $U$. The top panels show the behaviour of the parameters of the cathode and anode $Ag$ layers, the bottom ones refer to the $A{l_2}{O_3}$ layers.}
\end{figure*}

	The voltage dependence of the optical parameters of MDM layers are presented in Figure~\ref{fig:fig06}. It is seen that all the optical parameters of the $Ag$ layers experience noticeable changes at $U\sim8V$ и and $U\sim16V$. Similar irregular behaviour of the parameters is observed for the $A{l_2}{O_3}$ layers. Between these two voltage points parameters $n$ and $k$ keep fairly stable for all layers. The exclusion is the absorption coefficient of the $A{l_2}{O_3}$ layer which grows slightly on the interval  $U =20-30V$. It is interesting that the refractive index of the cathode $Ag$ layer falls to zero when $U > 16V$, which means that the dielectric permittivity becomes strictly negative and loses its imaginary component.
	
	Following the changes of parameters $n$ and $k$, the reflection coefficient also changes sharply.  The most pronounced changes in the MDM structure reflection occur when the voltage exceeds $16V$. An experiment on this MDM structure showed  that the voltage increase  up to $U =20-30V$ results in a $4.2$-times decrease of the reflection coefficient at ${\theta} = {42.9^0}$ and $5.4$-times increase at ${\theta} = {55.8^0}$. Calculations show that the contrast of the reflectivity can be increased dramatically by varying the layer thicknesses slightly. For instance, if we increase the anode layer thickness to $54\,nm$, we get a structure which at fixed angle ${\theta} = {60.5^0}$ may be considered as a switcher: with zero reflection at $U=0$ and  with $20\% $ reflectivity when $U$ increases to $30V$.

\section{Conclusions}
We have discovered significant changes of the dielectric permittivity of silver when a constant voltage of up to $30V$ is applied to a MDM-structure waveguide. It is shown that we can modulate the reflection coefficient widely by varying the voltage. Looking forward, the effect may be used in development of electrically controlled optical valves.

	We could not so far find an acceptable theory that can explain the abrupt change of optical parameters of the MDM structure. This will be the goal of later research.

\begin{acknowledgments} 
The work was supported by the project 2.1 of the RAS Presidium and RFBR grant 11-07-92470.
\end{acknowledgments}

\bibliography{my}

\end{document}